\documentclass[12pt]{article}
\usepackage{epsfig}

\newcommand\pubdate{\today}
\newcommand\hepnumber{hep-ex/0101016}

\def\csumb{CERN\\
1211 Geneva 23, Switzerland\\
Email: Stephan.Wynhoff\@cern.ch}

\def\Title#1{\begin{center} {\Large\bf #1 } \end{center}}
\def\Author#1{\begin{center}{ \sc #1} \end{center}}
\def\Address#1{\begin{center}{ \it #1} \end{center}}

\newcommand\pubblock{\rightline{\begin{tabular}{l} 
         \pubdate\\ \hepnumber \end{tabular}}}
\newenvironment{Abstract}{\begin{quotation}  }{\end{quotation}}
\newenvironment{Presented}{\begin{quotation} \begin{center} 
             Presented at the\end{center}
      \begin{center}\begin{large}}{\end{large}\end{center} \end{quotation}}
\def\Acknowledgments{\bigskip  \bigskip \begin{center}
          \large\bf Acknowledgments\end{center}}

\makeatletter
\def\section{\@startsection{section}{0}{\z@}{5.5ex plus .5ex minus
 1.5ex}{2.3ex plus .2ex}{\large\bf}}
\def\subsection{\@startsection{subsection}{1}{\z@}{3.5ex plus .5ex minus
 1.5ex}{1.3ex plus .2ex}{\normalsize\bf}}
\def\subsubsection{\@startsection{subsubsection}{2}{\z@}{-3.5ex plus
-1ex minus  -.2ex}{2.3ex plus .2ex}{\normalsize\sl}}
\def\tableline{\noalign{
\hrule height.7pt depth0pt\vskip3pt}}

\renewcommand{\@makecaption}[2]{%
   \vskip 10pt
   \setbox\@tempboxa\hbox{\small #1: #2}
   \ifdim \wd\@tempboxa >\hsize     
       \small #1: #2\par          
     \else                        
       \hbox to\hsize{\hfil\box\@tempboxa\hfil}
   \fi}

 \def\citenum#1{{\def\@cite##1##2{##1}\cite{#1}}}
 
\newcount\@tempcntc
\def\@citex[#1]#2{\if@filesw\immediate\write\@auxout{\string\citation{#2}}\fi
  \@tempcnta\z@\@tempcntb\m@ne\def\@citea{}\@cite{\@for\@citeb:=#2\do
    {\@ifundefined
       {b@\@citeb}{\@citeo\@tempcntb\m@ne\@citea\def\@citea{,}{\bf ?}\@warning
       {Citation `\@citeb' on page \thepage \space undefined}}%
    {\setbox\z@\hbox{\global\@tempcntc0\csname b@\@citeb\endcsname\relax}%
     \ifnum\@tempcntc=\z@ \@citeo\@tempcntb\m@ne
       \@citea\def\@citea{,}\hbox{\csname b@\@citeb\endcsname}%
     \else
      \advance\@tempcntb\@ne
      \ifnum\@tempcntb=\@tempcntc
      \else\advance\@tempcntb\m@ne\@citeo
      \@tempcnta\@tempcntc\@tempcntb\@tempcntc\fi\fi}}\@citeo}{#1}}
\def\@citeo{\ifnum\@tempcnta>\@tempcntb\else\@citea\def\@citea{,}%
  \ifnum\@tempcnta=\@tempcntb\the\@tempcnta\else
  {\advance\@tempcnta\@ne\ifnum\@tempcnta=\@tempcntb \else\def\@citea{--}\fi
    \advance\@tempcnta\m@ne\the\@tempcnta\@citea\the\@tempcntb}\fi\fi}
\makeatother

\def\beq{\begin{equation}}
\def\eeq#1{\label{#1}\end{equation}}
\def\eeqn{\end{equation}}

\newenvironment{Eqnarray}%
   {\arraycolsep 0.14em\begin{eqnarray}}{\end{eqnarray}}
\def\beqa{\begin{Eqnarray}}
\def\eeqa#1{\label{#1}\end{Eqnarray}}
\def\eeqan{\end{Eqnarray}}

\let\bar=\overbar

\def\O{{\cal O}}

\def\Dslash{\not{\hbox{\kern-4pt $D$}}}
\def\dslash{\not{\hbox{\kern-2pt $\del$}}}

\def\msb{{\bar{\ssstyle M \kern -1pt S}}}

\def\lsim{\mathrel{\raise.3ex\hbox{$<$\kern-.75em\lower1ex\hbox{$\sim$}}}}
\def\gsim{\mathrel{\raise.3ex\hbox{$>$\kern-.75em\lower1ex\hbox{$\sim$}}}}

\newcommand{\EE}{\ensuremath{\mathrm{e^+e^-}}}
\newcommand{\FF}{\ensuremath{\mathrm{f\bar{f}}}}
\newcommand{\MM}{\ensuremath{\mathrm{\mu^+\mu^-}}}
\newcommand{\TT}{\ensuremath{\mathrm{\tau^+\tau^-}}}
\newcommand{\WW}{\ensuremath{\mathrm{W^+W^-}}}
\newcommand{\ZZ}{\ensuremath{\mathrm{ZZ}}}
\newcommand{\QQQQ}{\ensuremath{\mathrm{q q q q}}}
\newcommand{\QQLN}{\ensuremath{{\mathrm{q q} \ell \nu}}}

\newcommand{\XWW}{\ensuremath{\sigma_\mathrm{WW}}}

\def\GeV{\ifmmode {\mathrm{\ Ge\kern -0.1em V\,}}\else
                   \textrm{Ge\kern -0.1em V\,}\fi}%
\def\MeV{\ifmmode {\mathrm{\ Me\kern -0.1em V\,}}\else
                   \textrm{Me\kern -0.1em V\,}\fi}%
\newcommand{\Ss}{\ensuremath{\sqrt{\mathit{s}}}}
\newcommand{\Ssp}{\ensuremath{\sqrt{\mathit{s^\prime}}}}
\newcommand{\rf}{\mathrm{r}_f}
\newcommand{\jf}{\mathrm{j}_f}
\newcommand{\gf}{\mathrm{g}_f}

\newcommand{\rtoth}{\mathrm{r^{tot}_{had}}}

\newcommand{\jtoth}{\mathrm{j^{tot}_{had}}}

\newcommand{\gtoth}{\mathrm{g^{tot}_{had}}}

\newcommand{\MZ}{\ensuremath{\mathrm{M}_\mathrm{Z}}}
\newcommand{\GZ}{\ensuremath{\Gamma_\mathrm{Z}}}
\newcommand{\MZbar}{\ensuremath{\overline{m}_{\mathrm{Z}}}}
\newcommand{\GZbar}{\ensuremath{\overline{\Gamma}_{\mathrm{Z}}}}
\newcommand{\MW}{\ensuremath{\mathrm{M}_\mathrm{W}}}
\newcommand{\GW}{\ensuremath{\Gamma_\mathrm{W}}}
\newcommand{\MH}{\ensuremath{\mathrm{M}_\mathrm{H}}}

\begin{document}
\begin{titlepage}
\pubblock

\vfill
\def\thefootnote{\fnsymbol{footnote}}
\Title{Standard Model Physics Results from LEP2}
\vfill
\Author{Stephan Wynhoff}
\Address{\csumb}
\vfill
\begin{Abstract}
  At LEP2 many Standard Model predictions are tested up to centre-of-mass
  energies of 209~\GeV. Fermion pair production cross sections and asymmetries
  agree well with the theoretical expectation over the entire energy range. The
  measurements are used to determine the $\gamma$/Z interference and to search
  for contact interactions up to 20~TeV. The cross sections for single-W, ZZ and
  \WW\ production agree well with the expectations. The branching fractions of
  the W boson into hadrons and leptons are determined as well as the CKM matrix
  element $\vert V_{cs}\vert$.  Precise measurements of the W mass and width are
  presented yielding $\MW=80.427\pm0.046\GeV$ and $\GW=2.12\pm0.11\GeV$. All
  electroweak data are very consistent with the Standard Model predictions. In a
  combined fit using the recent value of
  $\Delta\alpha_\mathrm{had}^\mathrm{(5)}(\mathrm{s})$ the mass of the Higgs
  boson is constrained to \MH=$88^{+60}_{-37}$ \GeV.
\end{Abstract}
\vfill
\begin{Presented}
5th International Symposium on Radiative Corrections \\ 
(RADCOR--2000) \\[4pt]
Carmel CA, USA, 11--15 September, 2000
\end{Presented}
\vfill
\end{titlepage}
\def\thefootnote{\arabic{footnote}}
\setcounter{footnote}{0}

\section{Introduction}

The LEP accelerator has provided since its start in 1989 many possibilities to
check Standard Model \cite{standard_model,Veltman_SM} (SM) predictions. During
the first years the accelerator was operated at the Z-pole (LEP1) and the four
LEP experiments, ALEPH, DELPHI, L3 and OPAL collected some 15 million hadronic
and 2 million leptonic Z decays. These data allowed a precise determination of
the properties of the Z boson \cite{ewwg_ls2000-01}. In the second phase of LEP,
LEP2, the centre-of-mass energy, \Ss, was successively increased up to $\Ss =
209$~\GeV allowing the production of \WW~and \ZZ~pairs. More than 8000 W-pair
events have been collected per experiment and are used to determine in
particular the mass and width of the W boson~\cite{ewwg-ww00-01}. Combining the
LEP results with other electroweak precision measurements allows thorough
consistency tests of the SM and to constrain the mass of the Higgs
boson~\cite{gurtu,pietrzyk}.

To match the statistical accuracy of the large data samples collected at LEP --
especially at energies above the Z-pole -- the corresponding theory programs
have been improved. For 2-fermion processes the programs
ZFITTER~\cite{zfitter}, TOPAZ0~\cite{topaz0} and KKMC~\cite{kkmc} have now a
precision better than 0.2\% for the total hadronic and leptonic\footnote{For
  Bhabha scattering the precision is estimated to 2\% for an angular range of
  $30^\circ < \vartheta < 150^\circ$.}  cross sections at high energies. The
KKMC program covers the entire energy range from $\tau$-- and $b$--factories
over LEP to linear colliders. Also for 4-fermion processes adequate precision
has been reached.  Using the double-pole approximation~\cite{doublepole}
RacoonWW~\cite{racoonww} and YFSWW3~\cite{yfsww} calculate the \WW~cross section
within 0.4\% above the production threshold. The cross section for the process
$\EE \rightarrow W e \nu$ is calculated within 4-5\% accuracy by
WPHACT~\cite{wphact}, grc4f~\cite{grace} and WTO~\cite{wto} using the fermion
loop scheme~\cite{floop}. The programs YFSZZ~\cite{yfszz} and ZZTO~\cite{zzto}
predict the Z-pair production cross section within 2\%. Details can be found in
the proceedings of the LEP2MC workshop~\cite{lep2mc2f,lep2mc4f}. Generally there
is now an excellent match in precision between theoretical predictions and
experimental measurements.

\section{Fermion Pair Production}
At centre-of-mass energies well above the Z-pole photon radiation becomes
important. The effects to consider are initial and final state photon radiation,
interference between these and the production of additional fermion pairs by a
photon or Z boson. The main interest is in events where the annihilation took
place at a high effective centre-of-mass energy, \Ssp, which is defined as the
mass of the outgoing lepton pair or of the $\gamma^*/Z$ propagator. Results are
given by all four experiments for events with $\Ssp > 0.85\cdot\Ss$. The results
for the reactions $\EE \rightarrow \mathrm{hadrons} (\gamma)$, $\EE \rightarrow
\MM(\gamma)$ and $\EE \rightarrow \TT(\gamma)$ are combined taking properly into
account the statistical and systematical uncertainties and their
correlations~\cite{lep2ffbar}.

\begin{figure}[ht]
\begin{center}
\epsfig{file=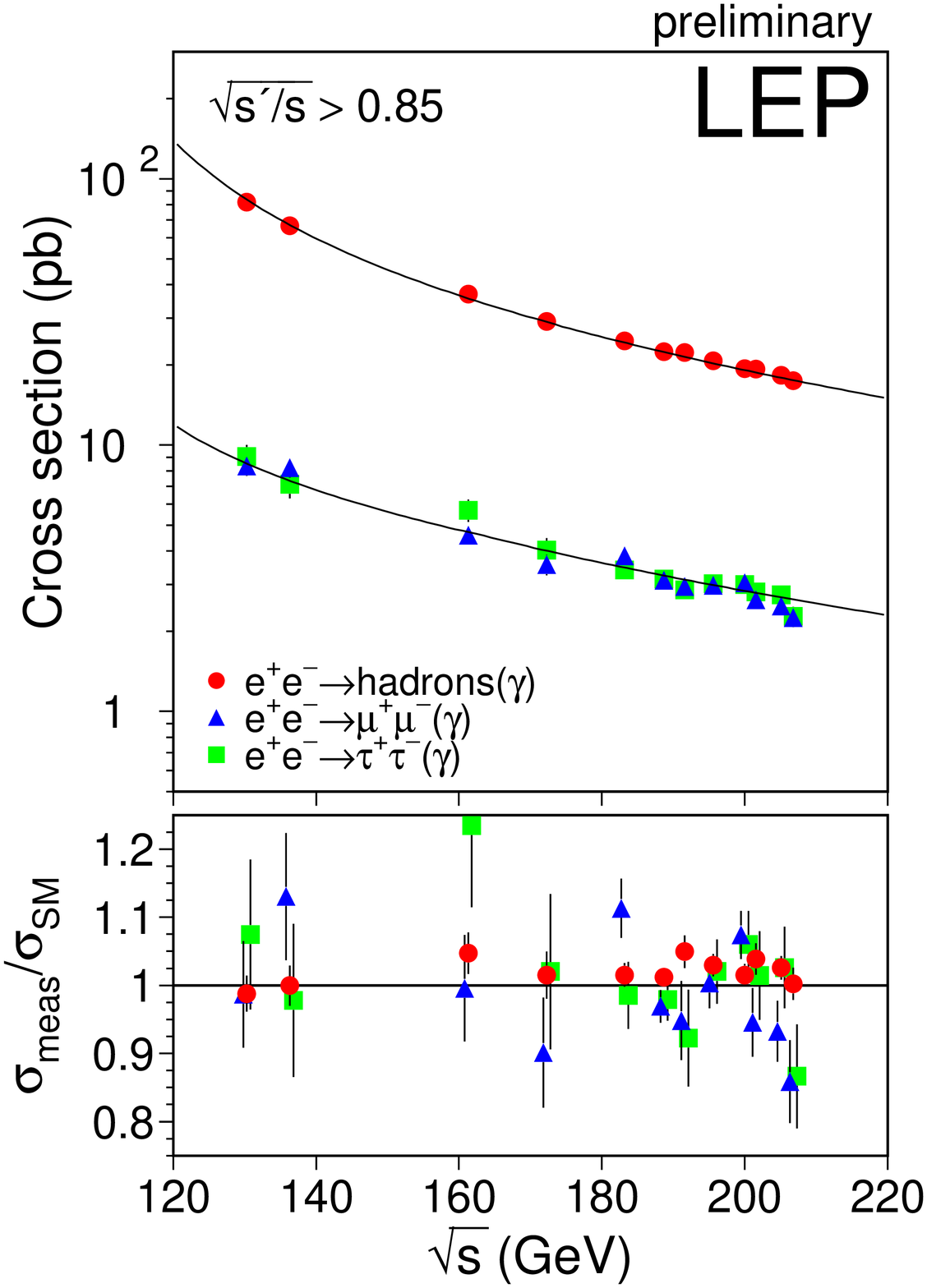,width=6.1cm}\hfill
\epsfig{file=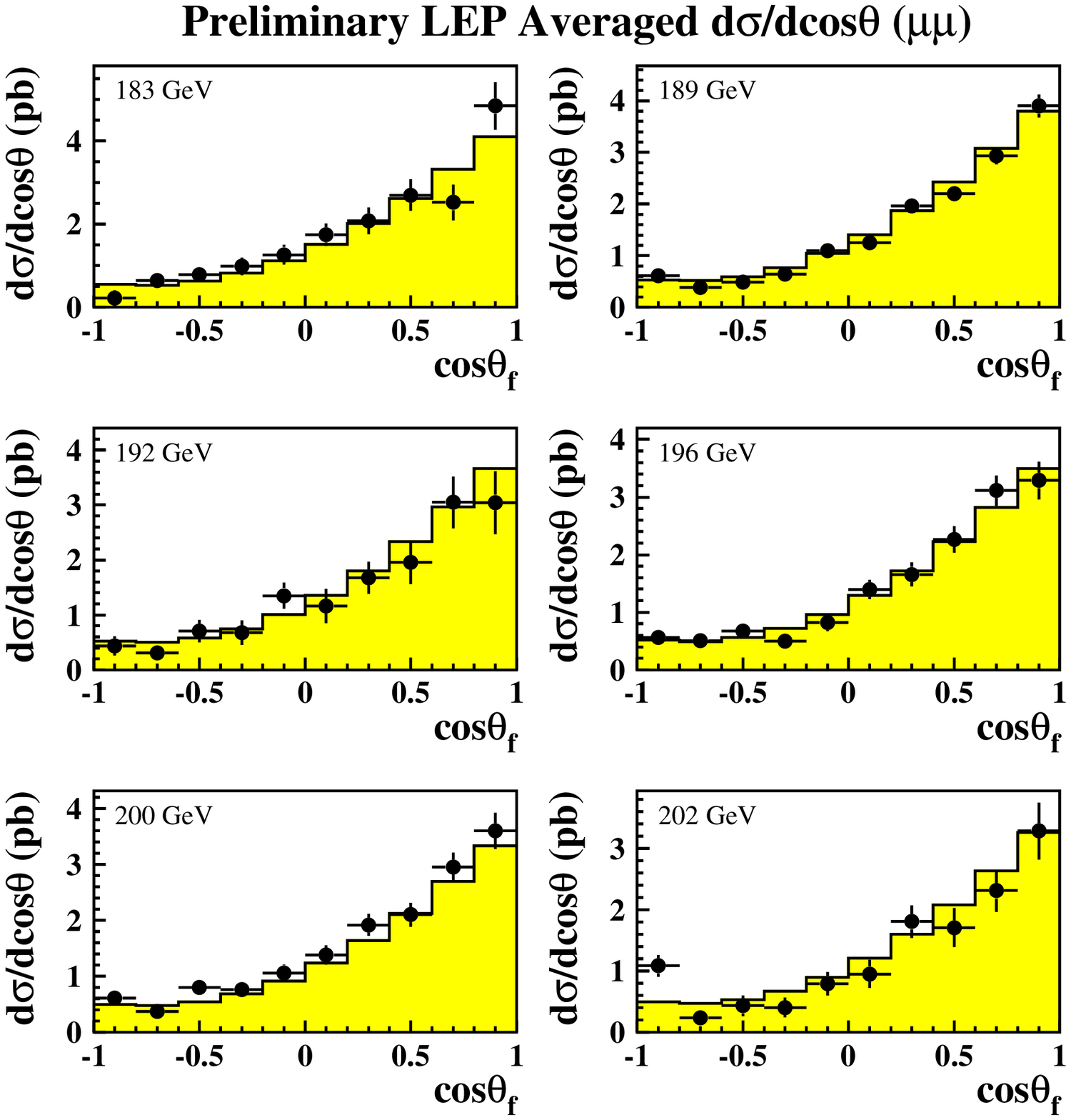,width=8.8cm}
\caption[0]{\label{mhtanb} The measured cross sections of fermion pair
  production and the differential cross section for muon pair production
  at LEP2.}
\label{fig:fppxs}
\end{center}
\end{figure}

The combined results for the total cross sections are in Figure~\ref{fig:fppxs}
compared to the SM predictions for all three processes. The measurements agree
well with the theoretical expectations.  For muon and tau pair production also
the differential cross sections, $d\sigma/d\cos(\vartheta)$, have been
determined.  The result for muon pair production is also shown in
Figure~\ref{fig:fppxs} for centre-of-mass energies from 183~\GeV\,to 202~\GeV.
Also the forward-backward asymmetries for these processes are in good agreement
with the SM.  For hadronic final states the ratios of cross sections for b
quarks and c quarks to the total hadronic cross section, $R_b$ and $R_c$, as
well as the forward-backward asymmetries for these flavours are determined.
Within the limited statistics of the measurements good agreement with the SM is
observed.

The reaction $\EE \rightarrow \FF$ has contributions from photon exchange, from
Z boson exchange and from $\gamma/Z$ interference. Within the S-Matrix
approach~\cite{smatrix} the lowest--order total cross sections and
forward--backward asymmetries are parametrised in the following way:
\begin{eqnarray*}
\sigma^0_{a}(s) & = &
  \frac{4}{3}\pi\alpha^2
                         \left[
                               \frac{ \gf^a}{s} +
                               \frac{{ \jf^a} (s-\MZbar^2) +{ \rf^a} \, s}
                                     {(s-\MZbar^2)^2 + \MZbar^2 \GZbar^2}
                         \right] ,
              \mathrm{for}~a=\mathrm{tot,fb} , \nonumber\\
&&\\
A^0_{\mathrm{fb}}(s) & = &
  \frac{3}{4} \frac{\sigma^0_{\mathrm{fb}}(s)}{\sigma^0_{\mathrm{tot}}(s)} ,
  \qquad\mathrm{with}\qquad\sigma^0_{fb}=\frac{4}{3}\left(\sigma_f-\sigma_b\right).
\end{eqnarray*}
The S--Matrix ansatz defines the Z resonance using a Breit--Wigner denominator
with an $s$--independent width. In other approaches, a Breit--Wigner denominator
with an $s$--dependent width is used, which implies the following transformation
of the values of the Z boson mass and width: $\MZ = \MZbar+34.1~\MeV$ and $\GZ =
\GZbar+0.9~\MeV$. In the following, the fit results are quoted after applying
these transformations. The S--Matrix parameters $\rf$, $\jf$ and $\gf$ give the
Z exchange, $\gamma$/Z interference and photon exchange contributions for
fermions of type $f$, respectively. For hadronic final states the parameters
$\rtoth$, $\jtoth$ and $\gtoth$ are sums over all produced quark flavours.

\begin{figure}[ht]
\begin{center}
\begin{minipage}[!]{7cm}
\epsfig{file=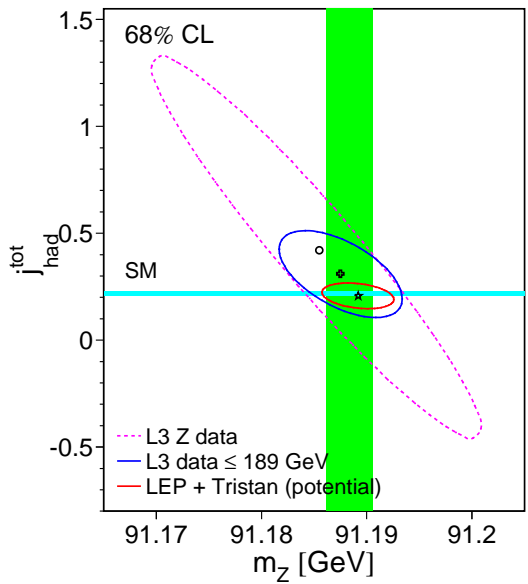,width=7cm,
  bbllx=0, bblly=0, bburx=470, bbury=540
}
\label{fig:mzjhad}
\end{minipage}
\hfill
\begin{minipage}[!]{7cm}
\caption{ 
  Contours in the ($\MZ$, $\jtoth$) plane at 68\% confidence level under the
  assumption of lepton universality. The dashed line is obtained from Z data
  only; the inclusion of 130~{\GeV} to 189~{\GeV} data gives the solid line.
  The circle (Z data) and the cross (all data) indicate the central values of
  the fits. The {SM} prediction for $\jtoth$ is shown as the horizontal band.
  The vertical band corresponds to the 68\% confidence level interval on {\MZ}
  in a fit assuming the Standard Model value for $\gamma$/Z interference. The
  smallest contour shows the result of a fit to all LEP and TRISTAN
  data.}
\end{minipage}
\end{center}
\end{figure}

\vspace*{-1.0cm}While in the standard fits to determine the Z boson mass the
$\gamma$/Z interference is fixed to its SM expectation in S-matrix fits it is
left free leading to an additional uncertainty on \MZ. Figure~\ref{fig:mzjhad}
shows the 68\% confidence level contours in the ($\MZ$, $\jtoth$) plane for the
L3 data taken at the Z--pole and after including the 130--189~{\GeV}
measurements~\cite{smatrixl3}. The improvement resulting from the inclusion of
the high energy measurements is clearly visible. The S-matrix fit agrees well
with the results from the standard fit indicated by the vertical band.
Figure~\ref{fig:mzjhad} also shows the potential~\footnote{Only some preliminary
  LEP1 results within the S-Matrix framework are available and
  systematic errors are not fully taken into account.} result when combining all LEP
data~\cite{lep2ffbar} and the Tristan~\cite{tristan} results. The total error on
\MZ\ is expected to be 2.3~\MeV\ showing that it is possible to remove the
additional uncertainty from the $\gamma$/Z interference on {\MZ} almost
completely.

The measured fermion pair cross sections and asymmetries can also be used to set
limits on contact interactions, fermion sizes, extra space dimensions, TeV
strings, gravitons and other new physics effects. For example, contact
interactions setting in at an energy scale $\Lambda$ can be described by the
following Lagrangian~\cite{cilang} where by convention the couplings $g$ are
normalised by $g^2/4\pi=1$ and the helicity amplitudes obey
$\vert\eta_{ij}\vert=0,1$:
\[
  {\cal L} = \frac{1}{1+\delta_{ef}}\;\;\;\sum_{i,j = \mathrm{L,R}} 
             \eta_{ij} 
             \frac{{ g}^2}{ \Lambda^2_{ij}}
             (\bar{\mathrm{e}}_i \gamma^{\mu}\mathrm{e}_i)
             (\bar{\mathit{f}}_j \gamma_{\mu}\mathit{f}_j),
\]
$\delta_{ef}$ is the Kronecker symbol being one for Bhabha scattering and zero
otherwise. 
\begin{table}[t!]
\caption{\label{tab:ci}
Preliminary limits on contact interactions from LEP combined data collected at
centre-of-mass energies from 130 GeV to 202 GeV.}
\begin{center}
\setlength{\tabcolsep}{9pt}
\renewcommand{\arraystretch}{1.2}
\begin{tabular}{lcccccc} 
\tableline
& \multicolumn{4}{c}{Helicity configuration} & 
\multicolumn{2}{c}{Energy scale [TeV]} \\ 
\cline{2-7}
\noalign{\vskip3pt}
 & $\eta_\mathrm{RR}$  & $\eta_\mathrm{LL}$  & $\eta_\mathrm{LR}$  &
 $\eta_\mathrm{RL}$ & $\Lambda_-$ & $\Lambda_+$ \\
\hline
AA & $\pm1$ & $\pm1$ & $\mp1$ & $\mp1$ & 13.9 & 17.6 \\
VV & $\pm1$ & $\pm1$ & $\pm1$ & $\pm1$ & 17.2 & 20.4 \\
RR & $\pm1$ & 0      & 0      & 0      &  9.7 & 12.3 \\
LL & 0      & $\pm1$ & 0      & 0      & 10.2 & 12.8 \\
\hline
\end{tabular}
\end{center}
\end{table}
A contact interaction, even at very high energy scales, can be detected at LEP2
by its interference effects with the SM by modifications to the differential
cross sections.
\[
  \frac{d\sigma}{d \cos\theta} =
  \frac{ d\sigma^\mathrm{SM}}{d \cos\theta} +
  c_{\mathrm{int}}(s,\cos \theta)\frac{1}{{\Lambda}^2} +
  c_{\mathrm{ci}}(s,\cos \theta)\frac{1}{{\Lambda}^4}.
\]
Such fits are done to the LEP combined measurements~\cite{lep2ffbar} and the resulting limits on
the energy scale are in the range from 10~TeV to 20~TeV depending on the
helicity configuration. The results are summarised in Table~\ref{tab:ci}.

\section{Boson Production Cross Sections}
The high centre-of-mass energies obtained at LEP2 allow the production not
only of fermion pairs but also of boson pairs, \WW and ZZ, and the production of
single W bosons.  

\begin{figure}[hbt]
\begin{center}
\epsfig{file=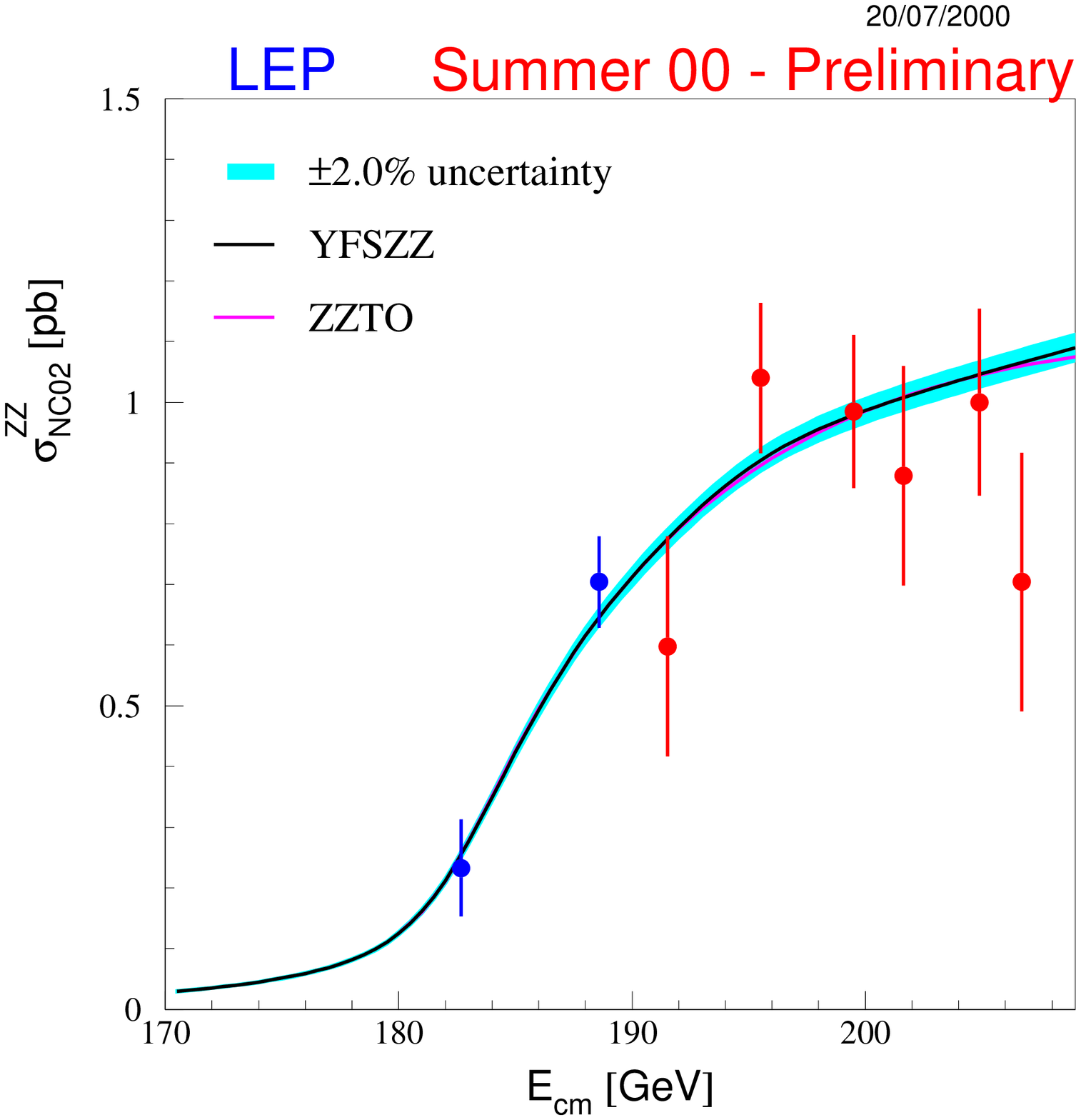,width=7.5cm}\hfill
\epsfig{file=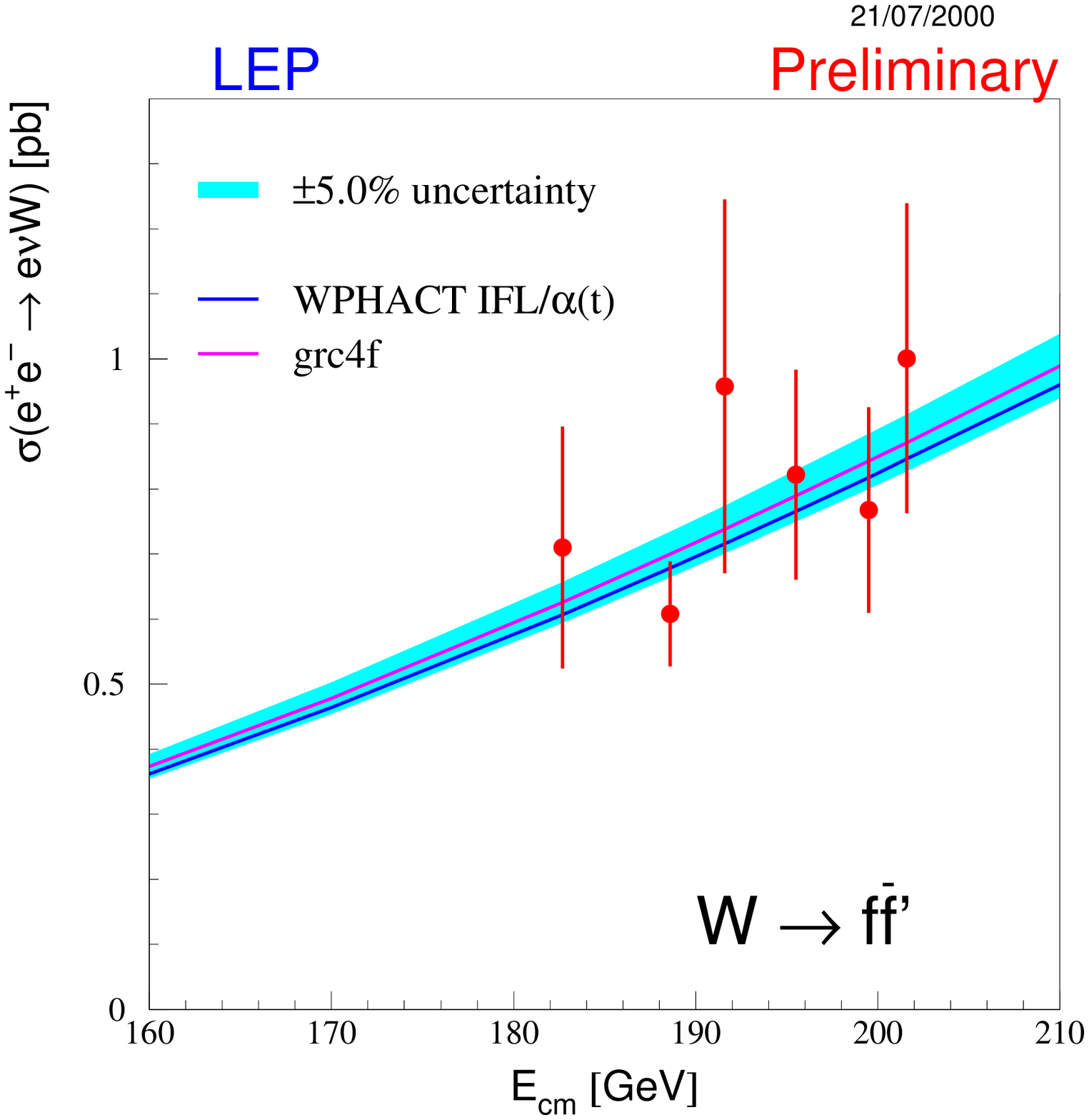,width=7.5cm}
\caption[0]{ The measured cross sections for Z pair production and
  for single W production.}
\label{fig:zzwen}
\end{center}
\end{figure}

The production of Z boson pairs tests the SM in the neutral-current sector and
is sensitive to scenarios for new physics like extra space dimensions or
couplings between neutral gauge bosons. All experiments have measured the ZZ
cross section at \Ss\ up to 208~GeV. The results are combined using the expected
statistical error and systematic uncertainties~\cite{lepxsboson}. They are
compared to predictions from YFSZZ and ZZTO in Figure~\ref{fig:zzwen} and show
no significant deviation from these theoretical models.

No new measurement for single W production ($\EE\rightarrow$We$\nu$) has been
provided above $\Ss = 202\GeV$ but the fermion loop scheme~\cite{floop} has been
introduced as an additional theoretical model. The data are compared with the
updated, slightly lower theoretical predictions in Figure~\ref{fig:zzwen}
showing good agreement.

\section{\WW Production}
At centre-of-mass energies above 160~\GeV\ the production of \WW\ pairs is
possible. Both W bosons decay into two fermions each producing three different
types of final states. About 45.6\% of the events decay fully hadronically.
These are balanced events of high multiplicity. In $3\times14.6$\% one W decays
hadronically while the other one decays leptonically resulting in 2 jets and a
high energetic lepton. A $\tau$ lepton can decay into a third, narrow jet
instead of an electron or muon. Fully leptonic decays are characterised by low
multiplicity and a lot of missing energy. The leptons are typically acoplanar.

\begin{figure}[hbt]
\begin{center}
\begin{minipage}[!]{7cm}
\epsfig{file=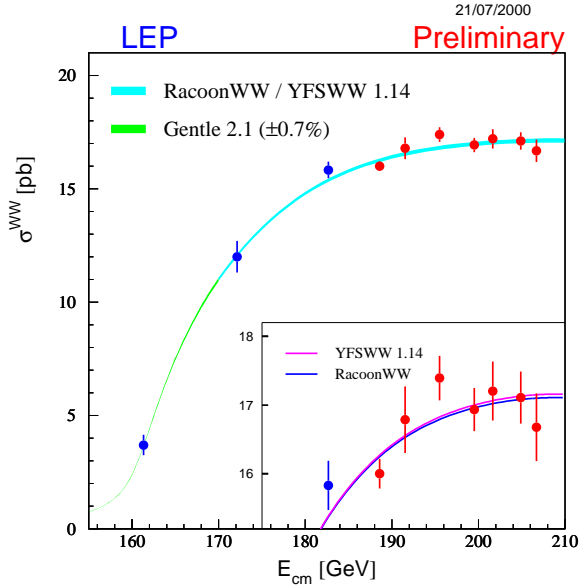,width=7cm,
  bbllx=0, bblly=0, bburx=470, bbury=540
}
\label{fig:xsww}
\end{minipage}
\hfill
\begin{minipage}[!]{7cm}
\caption{ 
 \WW production cross section at LEP. The points correspond to the combined
 measurements of all four LEP experiments for all decay channels. The lines
 represent the Standard Model predictions. The Gentle program is used for
 $\Ss<170\GeV$.}
\end{minipage}
\end{center}
\end{figure}
Events of all three topologies are selected by the four LEP experiments to
measure the total production cross section of \WW pairs. The combined LEP cross
section~\cite{lepxsboson} is shown in Figure~\ref{fig:xsww} and compared to the
predictions of the programs Gentle~2.1~\cite{gentle} (at centre-of-mass energies
below 170~\GeV) and RacoonWW and YFSWW~1.14 above threshold. Over the full
energy range an excellent agreement between the measurements and the SM is found.
\begin{table}[hbt]
\caption{\label{tab:wbr}
Preliminary hadronic and leptonic branching fractions of the W boson measured by the four
LEP experiments and the combined results. All numbers are given in percent.}
\begin{center}
\setlength{\tabcolsep}{9pt}
\renewcommand{\arraystretch}{1.2}
  \begin{tabular}{  l   r@{$ \, {\LARGE\pm} \, $}l r@{$ \, {\LARGE\pm} \, $}l
      r@{$ \, {\LARGE\pm} \, $}l r@{$ \, {\LARGE\pm} \, $}l  }
    \hline
     & \multicolumn{2}{c}{W$\rightarrow$hadrons} 
     &  \multicolumn{2}{c}{W$\rightarrow e\nu$}
     &  \multicolumn{2}{c}{W$\rightarrow \mu\nu$  }
       &  \multicolumn{2}{c}{W$\rightarrow \tau\nu$} \\
    \hline 
     ALEPH  & 67.22 & 0.53 & 11.19 & 0.34 & 11.05 & 0.32 & 10.53 & 0.42 \\
     DELPHI & 67.81 & 0.61 & 10.33 & 0.45 & 10.68 & 0.34 & 11.28 & 0.56 \\
     L3     & 68.47 & 0.59 & 10.22 & 0.36 &  9.87 & 0.38 & 11.64 & 0.51 \\
     OPAL   & 67.86 & 0.62 & 10.52 & 0.37 & 10.56 & 0.35 & 10.69 & 0.49 \\
    \hline        
     LEP    & 67.78 & 0.32 & 10.62 & 0.20 & 10.60 & 0.18 & 11.07 & 0.25 \\
    \hline   
  \end{tabular}
\end{center}
\end{table}

From the selected events also the decay fractions of the W boson into hadrons
and the three lepton flavours are determined. DELPHI and L3 used data from
centre-of-mass energies of 161~\GeV\ to 202~\GeV\ while ALEPH and OPAL analysed
data up to 207~\GeV. The results are listed in Table~\ref{tab:wbr}.  The
branching fractions for the three lepton flavours agree with each other and
support the hypothesis of lepton universality. The LEP combined leptonic
branching fraction of the W boson is $ Br(\mathrm{W}\rightarrow\mathrm{l}\nu) =
10.74\pm0.10$~\%. This direct measurement can be compared to the indirect
extraction at the TEVATRON where the combined results from CDF and
D0~\cite{tevatronbrwenu} yield $ Br(\mathrm{W}\rightarrow\mathrm{l}\nu) =
10.43\pm0.25$~\%.

From the hadronic branching fraction it is possible to determine the element
$\vert V_{cs}\vert$ of the Cabbibo-Kobayashi-Maskawa mixing matrix exploiting
the formula:
\[ \frac{Br(\mathrm{W}\rightarrow \mathrm{hadrons})}{1-Br(\mathrm{W}\rightarrow \mathrm{hadrons})} 
= \sum{ \left\vert V_{ij}^2\right\vert\left( 1+ \frac{\alpha_s}{\pi} \right)
  }.\] With LEP data a value of $\vert V_{cs}\vert = 0.989\pm0.016$ is obtained.
This value is in good agreement with the more direct determination using events
with tagged charm of $\vert V_{cs}\vert = 0.95\pm0.08$~\cite{lepxsboson}.

\section{W Mass Measurement}
The mass of the W boson is determined at LEP in two different ways. Close to the
production threshold the total cross section depends strongly on \MW. For \Ss =
161 - 172~\GeV\ the mass is determined from \XWW\ to be $\MW = 80.40\pm0.22 \GeV$
\cite{ewwg-ww00-01}. At higher centre-of-mass energies where the dependence of
\MW\ on \XWW\ is reduced the mass is reconstructed directly from the W decay
products.

\begin{table}[bht]
\caption{\label{tab:mwgw}
The values obtained for the mass and the width of the W boson obtained by the
four LEP experiments and their combination from data taken
at \Ss = 172 - 202 GeV. All numbers are preliminary.}
\begin{center}
\setlength{\tabcolsep}{9pt}
\renewcommand{\arraystretch}{1.2}
  \begin{tabular}{  l   r@{$ \, {\LARGE\pm} \, $}l 
r@{$ \, {\LARGE\pm} \, $}l }
    \hline
    & \multicolumn{2}{c}{\MW\ [\GeV]}     & \multicolumn{2}{c}{\GW\ [\GeV]} \\
    \cline{2-5}
    ALEPH  & 80.440 & 0.064 & 2.17 & 0.20 \\
    DELPHI & 80.380 & 0.071 & 2.09 & 0.15 \\
    L3     & 80.375 & 0.077 & 2.19 & 0.21 \\
    OPAL   & 80.485 & 0.065 & 2.04 & 0.18 \\
    \cline{2-5}
    LEP    & 80.427 & 0.046 & 2.12 & 0.11 \\
    \hline 
  \end{tabular}
\end{center}
\end{table}

From the three possible final states, qqqq, qql$\nu$ and l$\nu$l$\nu$, the fully
leptonic is not used because the two undetectable neutrinos inhibit the complete
determination of the event kinematics. For the other events leptons and jets are
reconstructed and \MW\ is determined in a kinematic fit to the measured fermion
energies and angles. Constraints from energy and momentum conservation -- one
for semileptonic and four for hadronic decays -- are imposed to improve the
resolution. In some analyses the two reconstructed W masses are required to be
equal as an additional constraint. For hadronic decays choosing the correct jet
pairing poses an additional problem. The pairing giving the best $\chi^2$ in the
fit is choosen. Possible gluon radiation is taken into account by splitting the
hadronic events into a 4- and 5-jet sample improving the mass resolution
(DELPHI, OPAL).

\begin{figure}[p]
\begin{center}
  \mbox{\epsfig{file=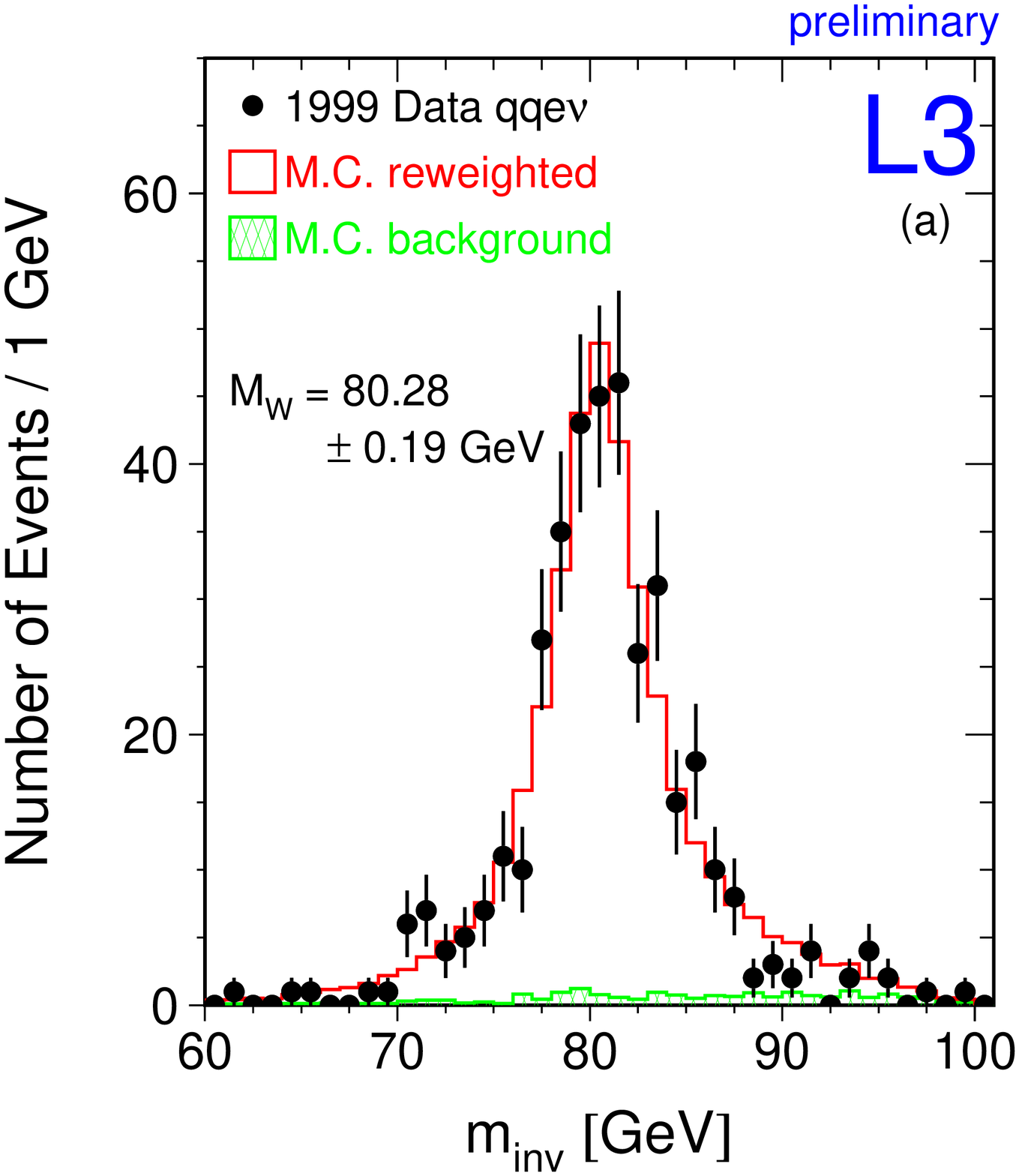,
      width=0.49\linewidth,height=0.35\textheight}}
  \mbox{\epsfig{file=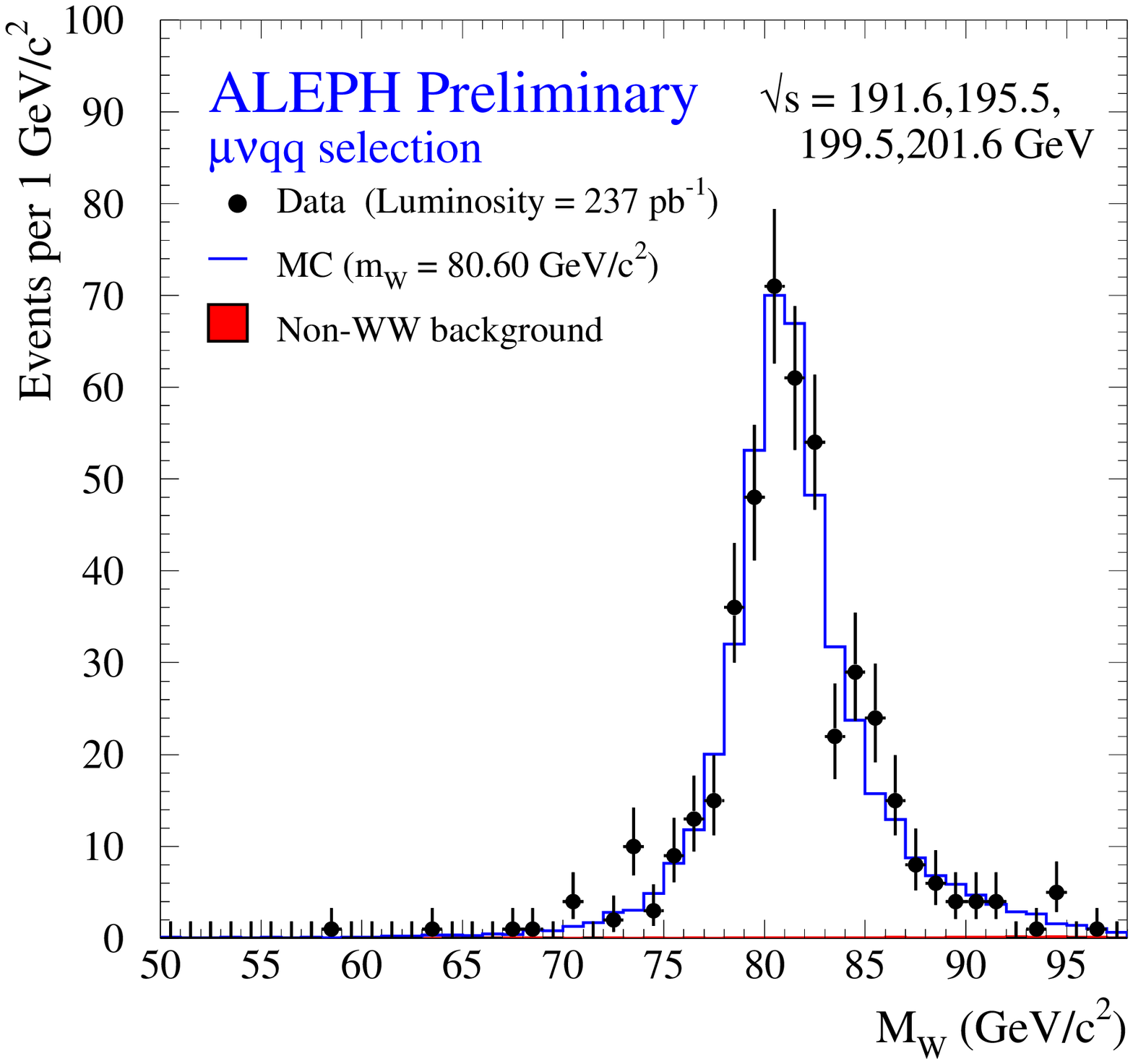,
      width=0.49\linewidth,height=0.35\textheight}}
  \mbox{\epsfig{file=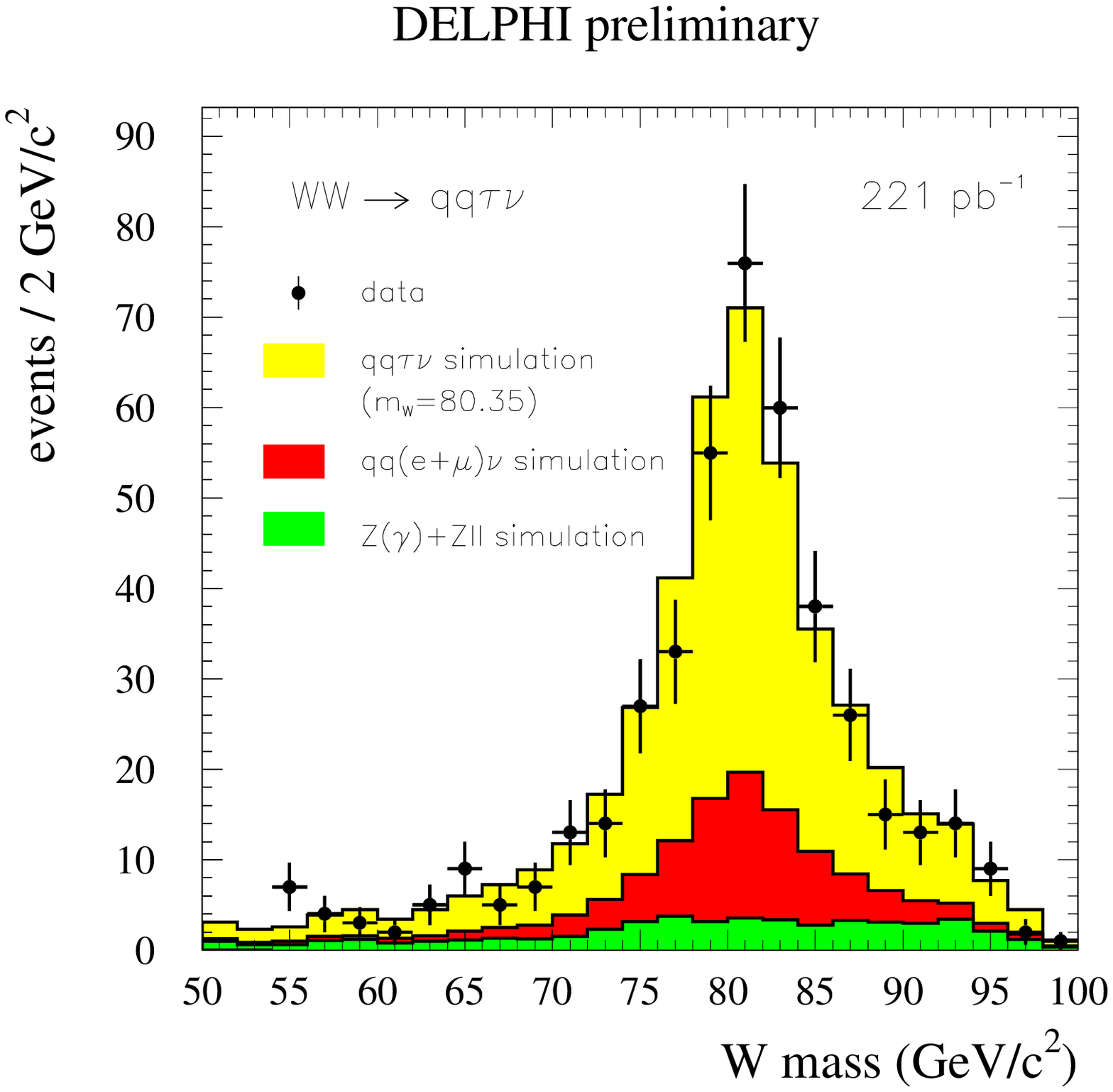,
      width=0.49\linewidth,height=0.35\textheight}}
  \mbox{\epsfig{file=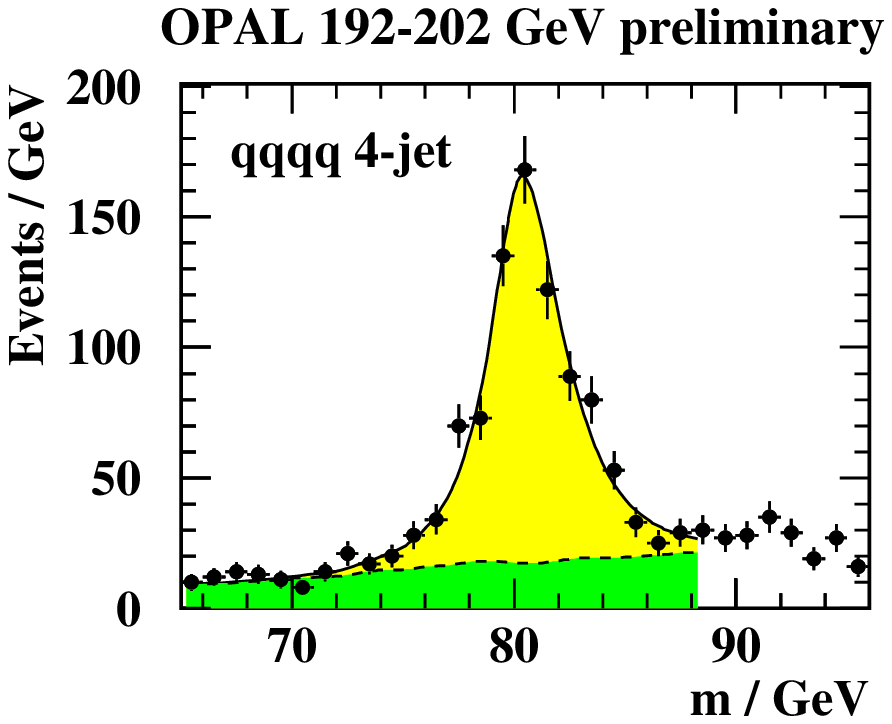,
      width=0.49\linewidth,height=0.35\textheight}}
\caption[0]{Invariant mass spectra for the four \WW\ production topologies used
      for direct reconstruction at $\Ss = 192 - 202$~GeV compared to the W mass
      fit results. L3, ALEPH and DELPHI use reweighted Monte-Carlo samples while
      OPAL took a relativistic Breit-Wigner function instead.}
\label{fig:mwspectrum}
\end{center}
\end{figure}

\begin{table}[hbt]
\caption{\label{tab:mwsys}
Breakdown of the systematic and statistical errors on \MW\ for the different
decay topologies in the LEP combined measurement.}
\begin{center}
\setlength{\tabcolsep}{9pt}
\renewcommand{\arraystretch}{1.2}
  \begin{tabular}{  l   r   r   r  }
    \hline
    Source & \multicolumn{3}{ c  }{Systematic Errors on \MW\ in \MeV} \\
    & \ \ \ \ \QQLN \ \ \ \ & \ \ \ \ \QQQQ \ \ \ \  & Combined \\
    \hline 
     Colour Reconnection &            --  &  50  &  13  \\
     Bose-Einstein Correlations &     --  &  25  &   7  \\
     LEP Beam Energy &                 17 &  17  &  17  \\
     ISR / FSR &                       8  &  10  &  8  \\
     Hadronisation &                   26 &  23  &  24  \\
    Detector Systematics &           11  &  8  & 10  \\
    Other &                           5  &  5  &  4  \\
    \hline        
     Total Systematic &                35  &  64  &  36  \\
    \hline   
     Statistical &                     38  &  34  &  30  \\
    \hline
     Total &                           51  &  73  &  47  \\
    \hline 
  \end{tabular}
\end{center}
\end{table}

The invariant mass distributions obtained from data taken at $\Ss = 192 -
202$~GeV are shown in Figure~\ref{fig:mwspectrum}. The W boson mass is extracted
from these spectra by comparing reweighted Monte-Carlo event samples 
corresponding to different mass hypotheses to data (ALEPH, L3, OPAL).
Alternativly, the differential cross sections are convoluted with resolution
functions (DELPHI, OPAL) or the mass is determined from a Breit-Wigner fit to
the measured mass spectrum (OPAL). 

The results for \MW\ are listed in Table~\ref{tab:mwgw}. The LEP value is a
combination of individual measurements performed at 172 - 202 GeV from the
experiments for different channels and years taking errors and correlations
into account. The resulting $\chi^2/\mathrm{dof}$ is $27.1/29$. The statistical
contribution to the error is 30~MeV, that from systematic uncertainties amounts
to 36~MeV.

Currently, the systematic uncertainties dominate the total error on \MW. A part
common to all measurements comes from the LEP beam energy determination and
amounts to 17~MeV at highest energies~\cite{lepebeam}. A new beam energy
spectrometer that has been installed in 1999 is expected to reduce this error to
7 - 12~MeV~\cite{lepspectrometer}. Other systematic uncertainties relevant for
all decay channels are hadronisation effects, detector related systematics and
effects of initial state and final state radiation.

The fully hadronic decays suffer from specific uncertainties due to hadronic
final state interactions (FSI). They occur because the distance between the two
decaying W bosons of about 0.1~fm is much smaller than the typical hadronic
interactions length of 1~fm. This can give rise to colour reconnection
effects~\cite{colrec} or Bose-Einstein correlations~\cite{be}.  Both can affect
the reconstruction of the invariant masses by momentum transfers between
particles that stem from different W bosons.  Combining the results from the
four experiments common uncertainties of 50~MeV for colour reconnection and
25~MeV for Bose-Einstein effects are estimated by comparing different
Monte-Carlo models. FSI effects may also show up in the difference between \MW\ 
values measured from semi-leptonic or from fully hadronic events. The
difference, determined removing systematic errors due to possible FSI effects,
amounts to
\[ \Delta\MW = \MW(\mathrm{qqqq}) - \MW(\mathrm{qql}\nu) = +5\pm50\MeV \]
and is compatible with zero. Recently, possible effects of FSI are also studied
in other observable than the W mass which are sensitive to
FSI~\cite{fsi-studies}, e.g. the particle flow in the overlap region between two
jets and particle correlation functions. In future it may be possible to exclude
some of the FSI models in a combined LEP analysis, which should reduce the
systematic uncertainty on $\MW$.

Table~\ref{tab:mwsys} shows a breakdown of all systematic errors for
semileptonic and hadronic final states. Due to the uncertainties related to FSI
effects the contribution of hadronic final states to the combined \MW\
measurement is only 27\% while the weight of the semileptonic events is 73\%.

The W boson mass $\MW = 80.427\pm0.046\GeV$ measured at LEP is in striking
agreement with its determination at $\mathrm{p\overline{p}}$
colliders~\cite{mwpp} of $\MW = 80.452\pm0.062\GeV$. The resulting average from
direct measurements is
\[ \MW = 80.436\pm0.037\GeV. \]
The method of direct reconstruction is also adequate to measure the width of the
W boson, \GW. The results of the four LEP experiments are shown in
Table~\ref{tab:mwgw}. The combination of the individual measurements is done in
the same way as for the determination of \MW. The resulting LEP value is
$\GW=2.12\pm0.11\GeV$ and is in agreement with the direct determination by
CDF~\cite{gwcdf} of $\GW=2.06\pm0.13\GeV$.

\section{Standard Model Fits}
Many SM parameters are measured at LEP1 and SLD like the mass and width of the Z
boson, \MZ\ and \GZ, the hadronic pole cross section, $\sigma^0_\mathrm{had}$,
the ratios of leptonic to hadronic widths, $R_\mathrm{l}$, the asymmetry
parameters for leptons and $b$ and $c-$ quarks, $A^\mathrm{0}_\mathrm{FB}$, the
$\tau$ polarisation and quark charge asymmetry, $Q_\mathrm{FB}$. At SLD
the measurement of left-right forward-backward asymmetry and recently the
asymmetry for $s$ quarks~\cite{sldafbs} are done. Finally, the result for the
on-shell value of $\sin^2\vartheta_\mathrm{W} = 0.2255\pm0.0021$ measured by
NuTEV/CCFR in $\nu$-nucleon scattering~\cite{nutev} and the value of
$\alpha(\MZ^2)$ are added. The latter can be expressed as
\[
  \alpha(\mathrm{s}) =  \displaystyle\frac{\alpha(0)}
  {1 - \Delta\alpha_\mathrm{lep}(\mathrm{s})
    - \Delta\alpha_\mathrm{had}^\mathrm{(5)}(\mathrm{s})
    - \Delta\alpha_\mathrm{had}^\mathrm{top}(\mathrm{s})}
\]
where all terms except the contribution from the five light quark flavours,
$\Delta\alpha_\mathrm{had}^\mathrm{(5)}(\mathrm{s})$, are know with high
accuracy. Here a value $\Delta\alpha_\mathrm{had}^\mathrm{(5)}(\mathrm{s}) =
0.02804\pm0.00065$~\cite{alphaold} is used. A fit within the Standard Model is performed to
these inputs to determine the parameters $\MZ, m_\mathrm{t}, \MH,
\alpha_\mathrm{s}$ and $\Delta\alpha_\mathrm{had}^\mathrm{(5)}(\mathrm{s})$.

\begin{figure}[bt]
\begin{center}
\epsfig{file=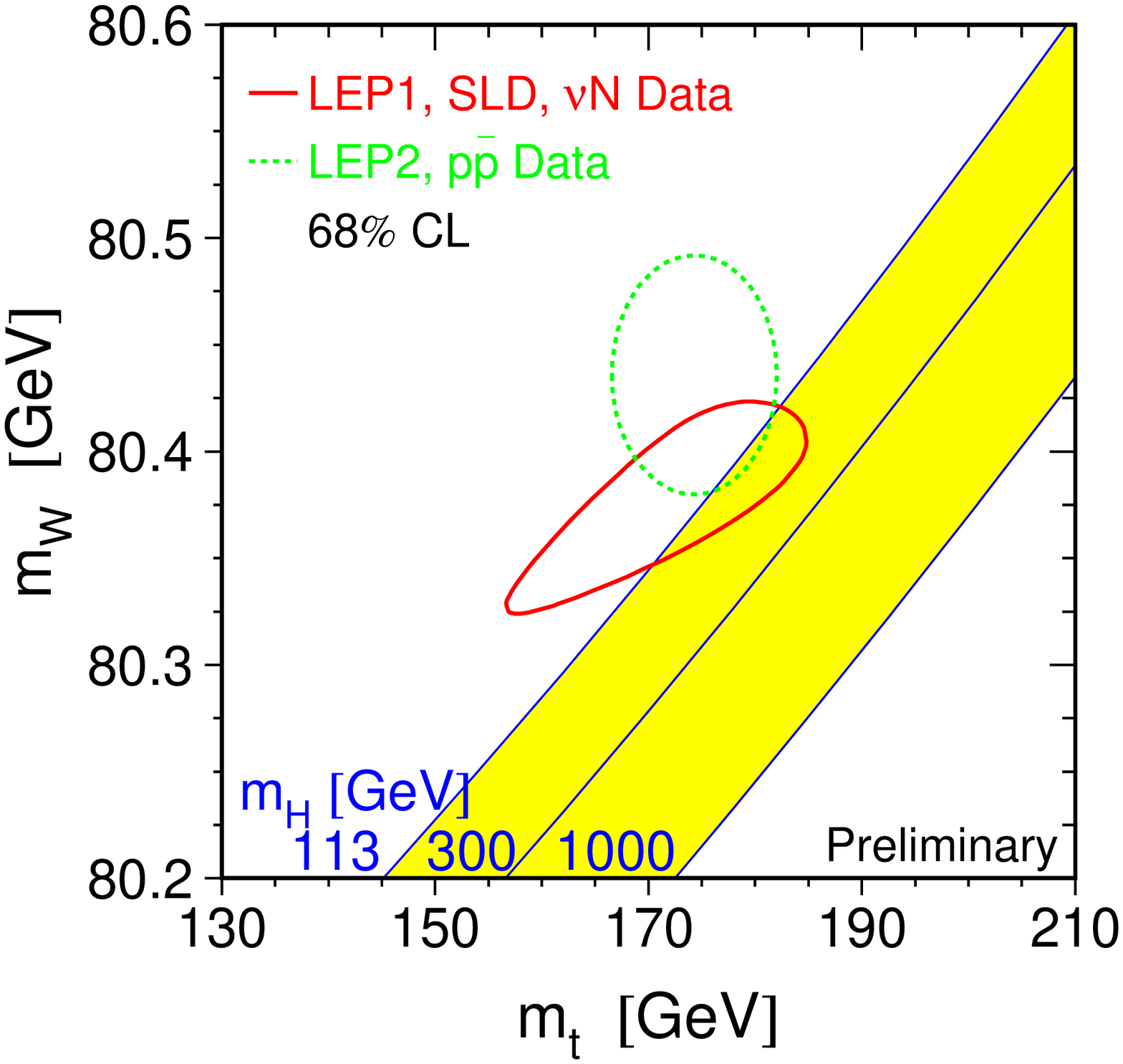,width=7.5cm}\hfill
\epsfig{file=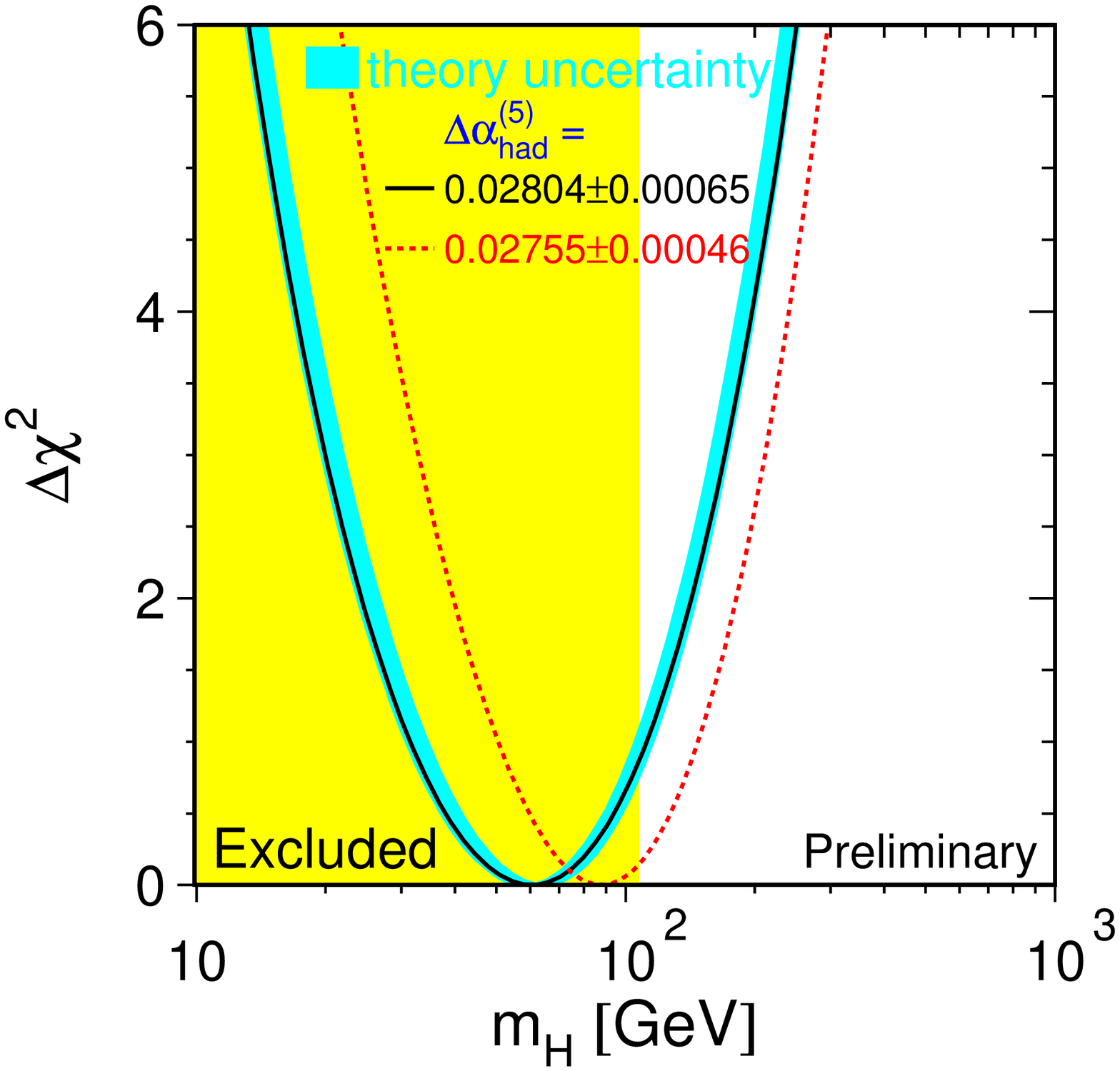,width=7.5cm}
\caption[0]{Left: Contours obtained from direct measurements and from a fit to
  electroweak precision data in the \MW-$m_\mathrm{t}$ plane testing the
  consistency of the SM. The results are compared to different values of \MH\\
  Right: $\Delta\chi^2$ of SM fits as a function of \MH for different values of
  $\Delta\alpha_\mathrm{had}^\mathrm{(5)}(\mathrm{s})$. The blue band indicates
  theoretical uncertainties due to higher order corrections. Higgs masses in the
  shaded area are excluded by direct searches.
}
\label{fig:smfit}
\end{center}
\end{figure}

In Figure~\ref{fig:smfit} the result of the fit is shown in the
\MW-$m_\mathrm{t}$ plane and compared to the direct measurements of \MW\ at
LEP and p$\overline{\mathrm{p}}$ colliders and of $m_\mathrm{t}$ at the
TEVATRON~\cite{tevatronmtop}. The measurements are nicely consistent with the
indirect determination from the SM fit.
A similar fit using all data except the direct measurement of the top quark mass
result in $m_\mathrm{t}=179^{+13}_{-10}\GeV$ and when using all data except
direct \MW\ determinations $\MW=30.386\pm0.025\GeV$ is obtained. Again, these
results are in good agreement with the respective direct measurements. This
demonstrates the compatibility and the internal consistency of the SM within the
existing precision and confirms the SM parameter relations at 1-loop level.

The SM fits can also be used to estimate the mass of the Higgs boson. To do this
a series of fits with fixed values of \MH\ is performed and the difference in
the $\chi^2$ values as shown in Figure~\ref{fig:smfit} is considered. Since the
leading radiative correction terms depend on $\log(\MH)$ the constraints that
can be obtained are not very stringent. The result using 
$\Delta\alpha_\mathrm{had}^\mathrm{(5)}(\mathrm{s})=0.02804\pm0.00065$ is
\[ \MH = 60^{+52}_{-29}\GeV;\;\;\;\log(\MH/\GeV)=1.78^{+0.27}_{-0.28}.\]
The slight decrease with respect to the previous result~\cite{lepew99} is mainly
caused by the change in \MW. The central value depends strongly on the top quark
mass and the value of $\Delta\alpha_\mathrm{had}^\mathrm{(5)}(\mathrm{s})$ used.

The value of $\Delta\alpha_\mathrm{had}^\mathrm{(5)}(\mathrm{s})$ is obtained by
integrating the $R_{had}$ distribution measured in \EE annihilation or
calculated in perturbative QCD:
\[ \Delta\alpha_\mathrm{had}^\mathrm{(5)}(\mathrm{s}) \propto  \displaystyle \int\limits_\mathrm{4 M_\pi^2}^\infty
\frac{\mathrm{R(s^\prime)\, ds^\prime}}{\mathrm{s^\prime(s^\prime - s)}}\]
Recent results obtained at BES~\cite{bes} have been used to extract the more
precise value
$\Delta\alpha_\mathrm{had}^\mathrm{(5)}(\mathrm{s})=0.02755\pm0.00046$~\cite{pietrzyk}
yielding a higher value for \MH:
\[ \MH = 88^{+60}_{-37}\GeV;\;\;\;\log(\MH/\GeV)=1.94^{+0.22}_{-0.24}\]
Relying on perturbative QCD the error on
$\Delta\alpha_\mathrm{had}^\mathrm{(5)}(\mathrm{s})$ is further reduced. With
the value
$\Delta\alpha_\mathrm{had}^\mathrm{(5)}(\mathrm{s})=0.02738\pm0.00020$~\cite{da5pqcd}
one obtains $\MH = 104^{+59}_{-39}\GeV$.

Depending on the value of $\Delta\alpha_\mathrm{had}^\mathrm{(5)}(\mathrm{s})$
used in the fit upper limits on the Higgs boson mass of 162 -- 215~GeV are
obtained at 95\% confidence level. The fits suggest that the Standard Model
Higgs is light. They are compatible with the results from direct searches for
the Higgs that exclude values of \MH\ below 113.5~GeV at 95\% C.L. and strongly
indicate the observation of a Higgs with a mass~\cite{lephiggs} of
\[ \MH=115^{+1.3}_{-0.9} \GeV. \]

\section{Conclusions}
Since its start in 1989 the energy range studied at LEP has more than doubled.
Up to \Ss = 209~\GeV the measurements of fermion pair production are in good
agreement with the Standard Model predictions. The data taken above the Z pole
allow to improve the determination of \MZ\ and the $\gamma$/Z interference within
the S-Matrix ansatz significantly. It also allows to exclude new (contact)
interactions below energy scales of 10~TeV to 20~TeV.

The cross sections for single W production, \WW and ZZ production agree with the
SM predictions as well.

From the large number of selected \WW pairs the mass and width of the W boson
can be directly reconstructed. The values
\begin{center}
  \begin{tabular}{l c r c l}
    \MW    & = & 80.427 & $\pm$ &  0.046 \GeV  \\
    \GW    & = & 2.12  & $\pm$ &  0.11 \GeV  \\
  \end{tabular}
\end{center}
are in perfect agreement with the indirect determination of these quantities in
fits to electroweak data. The impressive consistency between all direct
measurements and indirectly determined parameters confirms the Standard Model at
1-loop level. Fits to all electroweak data profit from the recent progress in the
determination of $\alpha(\MZ^2)$ and predict the mass of the Higgs boson to be
\begin{center}
  \begin{tabular}{l c r }
    \MH    & = & $88^{+60}_{-37}$ \GeV  \\
  \end{tabular}
\end{center}
which is consistent with the possible direct observation at LEP at \MH~ $\approx$ 115~\GeV.

\Acknowledgments
I thank the experiments ALEPH, DELPHI, L3, OPAL and SLD for making their most
recent and preliminary results available. I also thank the LEP and SLD working
groups on electroweak physics for their effort combining the measurements and
performing the fits. I am grateful to Arno Straessner for discussion of the W
mass measurement.

\end{document}